\documentclass{PoS}
\usepackage{amssymb,latexsym,amsmath}

\title{Uplifting Maximal Gauged Supergravities} 

\ShortTitle{Uplifting Maximal Gauged Supergravities} 

\author{\speaker{W. Baron}\\
         Universit\`a  degli studi di Padova and\\
				Universidad Nacional de La Plata\\ 
         E-mail: \email{wbaron@pd.infn.it, wbaron@fisica.unlp.edu.ar}} 

\abstract{
Which theories have a higher dimensional origin in String/M-theory is a non trivial question and it is still far from being understood in the constrained scenario of maximal supergravities. After 35 years of progress in this direction we have found supporting evidence in favor of the idea that every electric maximal supergravity in 4 dimensions can be uplifted to M-theory. We will review the current understanding of this problem with special emphasis in the uplifting of non compact supergravities and their relation with Exceptional Generalised Geometry.
}

\FullConference{18th International Conference From the Planck Scale 
to the Electroweak Scale \\            25-29 May 2015\\            Ioannina, Greece } 

\begin{document} 

\section{Introduction} 

Supergravities with maximal amount of supercharges in $4D$, {\it i.e.} $N=8$ have played a central role since the very discovery.
The reason is that the high degree of supersymmetry has deep implications making these theories simple but rich enough to describe very interesting phenomenological properties. In particular, these lead to metastable de Sitter solutions \cite{Dall'Agata:2012sx} and there is supporting evidence in favor that the Abelian $N=8$ supergravity could be a UV finite theory of gravity. In addition these are also relevant for applications in Condense Matter Theories via holography. 

The Abelian version was fully constructed in the seminal work of Cremmer and Julia \cite{Cremmer:1979up}, where they realized that the naive $GL(7,{\mathbb R})\times SO(7)$ symmetry is enhanced to global $E_7\times$ local $SU(8)$. This theory was obtained by dimensional reduction of 11D supergravity on an internal 7D torus and it has only one hypermultiplet containing a graviton $e^a{}_\mu$, 70 scalars parameterizing $E_7/SU(8)$, 28 vectors $A^a_\mu$, 8 gravitini $\psi^A_\mu$ and 56 spin $\frac12$ fields $\chi_{ABC}$. The first non Abelian deformation preserving $N=8$ was obtained by Scherk and Schwarz \cite{Scherk:1979zr} by compactification on twisted tori and leading to the so called flat groups, relevant at those times because they led to vanishing cosmological constants. Later, a different $N=8$ supergravity with an $SO(8)$ gauge group was proposed by de Wit and Nicolai \cite{deWit:1981eq} and later on similar theories with non compact gauge groups were constructed by analytic continuation by Hull \cite{Hull:1983yf}.

Another interesting set of deformations were proposed recently by Dall'Agata, Inverso and Trigiante \cite{DallAgata:2011aa} leading to an infinite number of $SO(8)$, $SO(p,q)$ and $CSO(p,q,r)$ theories where the gauge group is dyonically embedded as opposed to the original electric ones. 

Before the introduction of non Abelian couplings, there is an infinite family of equivalent lagrangians for maximal supergravity which can not be connected by local field redefinitions. These equivalent theories are determined by the way in which the duality group $E_7$ is embedded in the symplectic group $Sp(56,{\mathbb R})$ that mixes the 28 electric and 28 dual magnetic vector fields. 

The theory is fully determined once we chose the subgroup of the global $E_7$ to be gauged and after selecting which 28 out of the 56 vector fields will be considered as the physical vectors. All these possible deformations are parametrized by a unique object, the {\it embedding tensor} $\Theta_{\mathbb M}{}^\alpha\in {\bf 56+133}$ \cite{deWit:2007mt}. For instance, it appears in the covariant derivative $D_\mu=\partial_\mu-A^{\mathbb M}_\mu \Theta_{\mathbb M}{}^\alpha t_\alpha$, selecting the 28 gauge fields and picking at most 28 generators of $\mathfrak{e}_{7(7)}$. In addition, it must satisfy certain linear and quadratic constraints in order to ensure consistency with supersymmetry and locality.

As will be clear later, supersymmetry is the organizing principle in the uplift program and a special role is played by the supersymmetry transformations of the 28 electric and their 28 dual magnetic fields
\begin{eqnarray}
	\delta A_{\mu}^{AB} &=& \frac{1}{4 \sqrt2} (\Gamma^{IJ})_{AB} \left(u^{ij}{}_{IJ} + v^{ij\,IJ}\right)  \left(\overline \epsilon^k \gamma_\mu \chi_{ijk}+ 2 \sqrt2\, \overline \epsilon_i \psi_{\mu j}\right) + \rm{h.c.}\,, \label{susyvec}\\
	\delta A_{\mu\,AB} &=& -\frac{i}{4 \sqrt2}\,  (\Gamma^{IJ})_{AB} \left(u^{ij}{}_{IJ} - v^{ij\,IJ}\right) \left(\overline \epsilon^k \gamma_\mu \chi_{ijk}+ 2 \sqrt2\, \overline \epsilon_i \psi_{\mu j}\right) + \rm{h.c.}\,. \label{susydualvec}
\end{eqnarray}
where $A,B=1,\dots,8$ are $SL(8)$ indices and $u^{ij}{}_{IJ}=(u_{ij}{}^{IJ})^*$ and $v^{ij\;IJ}=(v_{ij\;IJ})^*$ denote the components of the scalar fields 
\begin{eqnarray}
{ \cal V}(x)=\left(\begin{matrix}u_{ij}{}^{IJ}&v_{ij\;IJ}\cr v^{ij\;IJ}&u^{ij}{}_{IJ}\end{matrix}\right).
\end{eqnarray}

Contrary to tori reductions, understanding the higher dimensional origin of other maximal supergravities like $SO(8)$ or $SO(p,q)$ was much more laborious and it was only very recently fully understood. The first attempts to interpret the $SO(8)$ gauged maximal supergravity as a Kaluza Klein (KK) on an internal seven sphere failed because the first (discarded) massive states in the KK tower transform into massless states after susy transformations leading to an inconsistent truncation. The solution to this issue appeared after replacing the linear KK perturbations by fully nonlinear expansions for the metric \cite{de Wit:1986iy}, 3-form \cite{deWit:2013ija}, dual 6-form potential \cite{Godazgar:2013dma} of the $SO(8)$ gauged supergravities and those for $SO(p,q)$ in \cite{Baron:2014bya}. So far these ansatze were checked with all known critical points of the potential of 4D supergravities with electric gaugings leading to known but also to new solutions of the field equations of 11D supergravity.

Similar to maximal supersymmetric theories in 4D, in 11D there is only one supermultiplet, which is composed by a graviton $E^{\hat A}{}_{\hat M}$, a  3-form potential $A_{\hat M\hat N\hat P}$ and a Rarita-Schwinger $\Psi_{\hat M}$. 
The bosonic fields transform under susy as 
\begin{eqnarray}
\delta E^{\hat A}{}_{\hat M}&=&-i\;\frac{\sqrt2}{2}\; \bar\epsilon \;\tilde\Gamma^{\hat A}\Psi_{\hat M}\,,\\
\delta A_{\hat M\hat N\hat P}&=&\frac32\; \bar\epsilon \;\tilde\Gamma_{[\hat M\hat N}\;\Psi_{\hat P]}\,.
\end{eqnarray}

In addition to these fields, sometimes it is useful to also consider the dual 6-form defined via 
\begin{eqnarray}
\partial_{[\hat M_1}\tilde A_{\hat M_2\dots \hat M_7]}=-\frac{1}{4! 7}\epsilon_{\hat M_1\dots \hat M_{11}}F^{\hat M_8\dots \hat M_{11}}-\frac{5\sqrt2}{2}A_{\hat M_1 \hat M_2 \hat M_3}F_{\hat M_4\dots \hat M_7}\,,\label{6form}
\end{eqnarray}
whose susy variation is given by 
\begin{eqnarray}
\delta \tilde A_{\hat M_1\dots \hat M_6}&=&-\frac{3}{\sqrt2}\; \bar\epsilon \;\tilde\Gamma_{[\hat M_1\dots \hat M_5}\;\Psi_{\hat M_6]}
+15\; \bar\epsilon \;\tilde\Gamma_{[\hat M_1\;\hat M_2}\;\Psi_{\hat M_3}A_{\hat M_4 \hat M_5 \hat M_6]}\,.
\end{eqnarray}

Compactifications on twisted tori are very well known since the seminal work of Scherk and Schwarz \cite{Scherk:1979zr}. In these reductions the uplift ansatz is quite simple. The internal components of the vielbein has a linear decomposition as the product of a 4D scalar matrix times a twist matrix depending on 7D internal coordinates, which can be interpreted as the vielbein of a background twisted tori. 

The situation is much more complicated if the internal space is not flat. For instance it took more than 30 years to fully understand the simplest not flat situations, $i.e.$ the reduction on a seven sphere. But very surprisingly it is possible to show that also truncations on spheres and hyperboloids can be interpreted as generalised Scherk-Schwarz reductions. 

In the next two sections we will study the uplifting problem from the two main available approaches in the literature, a bottom up and a top-down approach.

\section{Bottom up approach}

The starting point to get the full nonlinear uplift ansatze is the non canonical on-shell field redefinition of 11D supergravity in terms of fields covariant under $SU(8)$. 

The original formulation of 11D supergravity has an invariance under diffeomorphisms + local $SO(1,10)$ + Abelian gauge transformations + local $N=1$ SUSY. On the contrary, the $SU(8)$ reformulation is obtained by first fixing a triangular gauge for the vielbein breaking the local $SO(1,10)$ to $SO(1,3)\times SO(7)$ and then enlarging it to $SO(1,3)\times SU(8)$.

The KK ansatz for the vector gauge fields appearing in the 11-dimensional metric components, $B_\mu^m$, is that it is proportional to the Killing vectors corresponding to the isometries of the background space. Our proposal is that the same ansatz can be used also for some particular situations where there is no solution in the higher dimensional theory. To be more precise we want to focus here on the $SO(p,8-p)$ gauged supergravities. We claim that the correct KK ansatz for the vector fields is to factorize the internal coordinate dependence in terms of the Killing vectors of the seven dimensional metric whose isometries are $SO(p,8-p)$. Notice that such a space is the round 7D sphere for the $p=8$ case, but it is a non compact Lorentzian hyperboloid for $p=1,\dots,7$. As will be clear later, the ansatz still relates the critical points of 4D with Euclidean 7D internal solution of 11D supergravity. The 28 Killing vectors appearing in the ansatz are well defined on a Euclidean space, but only part of them can be interpreted as generating isometries of the internal space. The 4D consequence is the well known property that critical points of $SO(p,q)$ gauged supergravity break the gauge symmetry to a compact subgroup, being $SO(p)\times SO(q)$ the maximal allowed. Hence
\begin{equation}
	B_\mu{}^m(x,y) = -\frac12\, K^m{}_{AB}(y)\, A_\mu{}^{AB}(x),\label{Bmansatz}
\end{equation}
while the dual vector fields, appearing in the 3-form components $A_{\mu mn}$ are multiplied by the covariant derivatives of the Killing vectors, $K_{mn}{}^{AB} \equiv R^{-1}\,\stackrel{\circ}g_{L\,mp}\, \stackrel{\circ}{\nabla}_n K^{p AB}$, where $\stackrel{\circ}g_{L\,mn}$ is the (Lorentzian for $p\neq8$) metric of the maximally symmetric spaces
\begin{equation}
	B_{\mu mn}(x,y) = -\frac{1}{2\sqrt2}\,\, K_{mn}{}^{AB}(y)\, A_{\mu\,AB}(x),\label{Bmnansatz}
\end{equation}
where $B_{\mu mn}(x,y) \equiv A_{\mu mn} - B_\mu^p A_{mnp}$ and indices $A,B$ in the Killing tensors are raised and lowered with the Lorentzian flat metric with $(p,8-p)$ signature.

Other relevant fields, are the dual ones $B_{\mu m}$ and $B_\mu{}^{mn}$ which, in addition depend on the dual 6-form to get a fully covariant vector field ${\cal B}_\mu{}^{\mathbb M}=({\cal B}_\mu{}_{MN},{\cal B}_\mu{}^{MN})$ with ${\cal B}_\mu{}_{mn}=B_\mu{}_{mn}$, ${\cal B}_\mu{}_{m8}=B_\mu{}_{m}$, ${\cal B}_\mu{}^{mn}=B_\mu{}^{mn}$ and ${\cal B}_\mu{}^{m8}=B_\mu{}^{m}$. 

The $d.o.f.$ of the 11D gravitino $\Psi_M$ can also be rearranged in an $SU(8)$ covariant form. In fact, it is possible to introduce appropriate chiral transformations extending the $SO(7)$ tangent space group for Majorana spinors to $SU(8)$. Firstly, one splits 11 into 4+7, $\Psi_M\rightarrow \left(\Psi_\mu\,,\;\Psi_m\right)$ and performs some redefinitions to have the parity assignments of 4D\footnote{We use the representation of the 11D gamma matrices, $\gamma'$ in terms of the gamma's of 4D and 7D, $i.e.$ $\Gamma'_\alpha=\gamma_\alpha\otimes 1$, $\Gamma'_a=\gamma_5\otimes\Gamma_a$.
}
\begin{eqnarray}
\tilde\phi_\mu=\tilde e_\mu{}^\alpha \Delta^{-\frac14}(i\gamma_5)^{-\frac12}
\left(\Psi_\alpha-\frac12\gamma_5\gamma_\alpha\Gamma^a\Psi_a\right)\;,
\;\;\;\;\;\;\;\;\;\;\;\;\; \tilde\phi_a=\Delta^{-\frac14}(i\gamma_5)^{-\frac12} \Psi_a\;,\;\;
\end{eqnarray}
where we have introduced $\tilde g_{\mu\nu}=\Delta\, g_{\mu\nu}$ and $\Delta=\sqrt{|\det(g_{mn})\,\det(\stackrel{\circ}g_L{}^{pq})|}$, being $g_{\mu\nu}$ and $g_{mn}$ the 4D and 7D components of the 11D metric $g_{\hat M\hat N}$. Finally one promotes the spin-7 to chiral $SU(8)$\footnote{
Similarly, the appropriate redefinitions for the susy parameters are
\begin{eqnarray}
\hat\epsilon^A=\frac{1+\gamma_5}{2}\epsilon'_A\;,\;\;\;\;
\hat\epsilon_A=\frac{1-\gamma_5}{2}\epsilon'_A\;,\;\;\;\;
\epsilon'=\Delta^{\frac14}(i\gamma_5)^{-\frac12}\epsilon.
\end{eqnarray}} 
\begin{eqnarray}
\hat\phi_\mu{}^A&=&\frac{1+\gamma_5}{2} \phi'_\mu{}_A\;,
\;\;\;\;\;\;\;\;\;\;\;\;\;\;\;\;\;\;\;\;\;\;\;\;\;\;\;\;\;\;\hat\phi_\mu{}_A=\frac{1-\gamma_5}{2} \phi'_\mu{}_A\;,\cr
\hat\chi^{ABC}&=&\frac{1+\gamma_5}{2} \frac{3}{\sqrt2} i \Gamma^a_{[AB}\phi'_{a\;C]}\;,
\;\;\;\;\;\;\;\;\;\;\;\;\hat\chi_{ABC}=\frac{1-\gamma_5}{2} \frac{3}{\sqrt2} i \Gamma^a_{[AB}\phi'_{a\;C]}\;.
\end{eqnarray}

The 11D susy transformations of the vector fields now read
\begin{equation}
	\delta {\cal B}_\mu{}^{MN}(x,y) = \frac{\sqrt{2}}{8}\, \hat V^{AB\; MN}(x,y) \left(\overline \epsilon^C \gamma_\mu \chi_{ABC}+ 2 \sqrt2\, \overline \epsilon_A \psi_{\mu B}\right)(x,y) + \rm{h.c.}\,,\label{deltaBElec}
\end{equation}
\begin{equation}
	\delta {\cal B}_{\mu MN}(x,y) = \frac{\sqrt{2}}{8}\, \hat V^{AB}{}_{MN}(x,y) \left(\overline \epsilon^C \gamma_\mu \chi_{ABC}+ 2 \sqrt2\, \overline \epsilon_A \psi_{\mu B}\right)(x,y) + \rm{h.c.}\,.\label{deltaBMag}
\end{equation}
where we have introduced the coset representative $\hat V_{\mathbb A}{}^{\mathbb M}\in E_7/SU(8)$. Here we will focus only on $\hat V_{AB}{}^{m8}, \hat V_{AB}{}^{mn}$ whose explicit expressions are given by
\begin{eqnarray}
\hat V_{AB}{}^{m8}&=&i \Delta^{-\frac12} e_a{}^m \Gamma'{}^a_{CD}\Phi^C{}_A\Phi^D{}_B,\cr
\hat V_{AB}{}_{mn}&=&-\frac{1}{12} i\Delta^{-\frac12}\left( e^a{}_m e^b{}_n (\Phi^T \Gamma'_{ab}\Phi)_{AB}
+6\sqrt2 A_{mnp} (\Phi^T \Gamma'{}^p\Phi)_{AB}\right).\label{hatV}
\end{eqnarray}

Here we have introduced a local $SU(8)$ rotation, $\Phi$, in order to complete the $SU(8)$ covariance of the reformulation and we have also rotated the spinors $\chi^{ABC}=\Phi^A{}_{A'}\Phi^B{}_{B'}\Phi^C{}_{C'}\;\hat\chi^{A'B'C'}$, $\psi^A_\mu=\Phi^A{}_{A'}\;\hat\phi^{A'}_\mu$.

With the exception of the proposals in (\ref{Bmansatz}) and (\ref{Bmnansatz}), so far we have only introduced some field redefinitions and so this is just a rewriting of 11D supergravity. In order to perform a truncation to 4D we must supplement the ansatze for the vector fields with an ansatz for the fermionic fields. Again we propose a KK-like ansatz in terms of Killing spinors, $\eta_A{}^i$, $i.e.$
\begin{eqnarray}
\psi_{\mu\;A}(x,y)=\psi_{\mu\;i}(x)\eta_A{}^i(y)\;,\;\;\;\chi_{ABC}(x,y)=\chi_{ijk}(x)\eta_A{}^i(y)\eta_B{}^j(y)\eta_C{}^k(y)\;,\;\dots
\end{eqnarray} 

So, plugging the anstaze for the vectors and spinors with (\ref{deltaBElec}), (\ref{deltaBMag}) and comparing them with (\ref{susyvec}), (\ref{susydualvec}) lead to the identifications
\begin{eqnarray}
	V^m{}_{ij}(x,y) &\equiv& \hat V^m{}_{AB} \eta^A{}_i\eta^B{}_j 
	=\frac{\sqrt2}{16}\, K^m{}_{AB}(y) \,\Gamma'{}^{IJ}_{AB}\,(u_{kl}{}^{IJ}+v_{kl\,IJ})(x)\;,\;\;\;\;\;\; 
	V^{m\,ij} = (V^m{}_{ij})^*, \\[2mm]
	V_{mn\,ij}(x,y) &\equiv& \hat V_{mn\, AB} \eta^A{}_i\eta^B{}_j 
	= \frac{3\sqrt2}{8} i\, K_{mn}{}^{AB}(y) \,\Gamma'{}^{IJ}_{AB}\,(u_{kl}{}^{IJ}-v_{kl\,IJ})(x)\,,\,
	V_{mn}{}^{ij} = (V_{mn}{}_{ij})^*\,. \label{emnu} 
\end{eqnarray}

From these maps we can read the non linear ansatze for the internal components of the metric and the three form potential\newpage

\begin{eqnarray}
	\Delta^{-1} g^{mn} &=&4\, V^{m\,ij}V^n{}_{ij}\cr 
	&=&\frac{1}{32}\, K^m{}_{AB} \, K^n{}_{CD} \,(\Gamma^{IJ})_{AB} (\Gamma^{KL})_{CD}\, (u^{ij}{}_{IJ}+v^{ijIJ})\, (u_{ij}{}^{KL}+v_{ijKL}), \label{metricansatz}\\
	A_{mnp} &=& \frac{\sqrt2}{3} \, \Delta\, g_{pq}\, V_{mn\,ij} V^{q\,ij} \cr
	&=& -\frac{i}{32\sqrt2} \, \Delta\, g_{pq}\, K_{mn}{}^{AB}\,  K^q{}_{CD}\,(\Gamma^{IJ})_{AB} (\Gamma^{KL})_{CD}\, (u^{ij}{}_{IJ}-v^{ijIJ})\,(u_{ij}{}^{KL}+v_{ijKL})\,. \label{3formansatz}
\end{eqnarray}

These nonlinear ansatze are in perfect agreement with the literature. Certainly, all known critical points of 4D maximal $SO(p,q)$ supergravities with electric gaugings lead to a solution of the field equations of 11D supergravity if we assume a Freund-Rubin solution for the 4D components of the 4-form (see table \ref{solutions}).

\begin{table*}
	\centering
\begin{tabular}{||c | c | c| c||}
\hline
\hline
Gauge group & Remnant symmetry & Remnant SUSY & 4D space \\
\hline
$SO(8)$ &$SO(8)$ & $N=8$ & AdS\\
\hline
$SO(8)$ &$SO(7)_-$ & $N=0$ & AdS\\
\hline
$SO(8)$ &$SO(7)_+$ & $N=0$ & AdS\\
\hline
$SO(8)$ &$G2$ & $N=1$ & AdS\\
\hline
$SO(8)$ &$SU(4)$ & $N=0$ & AdS\\
\hline
$SO(8)$ &$SU(3)\times U(1)$ & $N=2$ & AdS\\
\hline
$SO(8)$ &$SO(3)\times SO(3)$ & $N=0$ & AdS\\
\hline
$SO(3,5)$ &$SO(3)\times SO(5)$ & $N=0$ & dS\\
\hline
$SO(4,4)$ &$SO(4)\times SO(4)$ & $N=0$ & dS\\
\hline
$SO(4,4)$ &$SO(3)\times SO(3)$ & $N=0$ & dS\\
\hline
\end{tabular}
	\caption{The non linear uplift ansatze was successfully tested with all the critical points of $SO(p,8-p)$ supergravity displayed above.}
	\label{solutions}
\end{table*}

\section{Exceptional Geometry interpretation}

The higher dimensional interpretation of 4D maximal supergravities can be described in a very natural and unified way in the framework of the so called {\it exceptional geometries}. These are U-duality covariant extensions of 11D supergravities like the Extended Geometry (EG) formalism \cite{EG} or Exceptional Field Theory (EFT) \cite{EFT} which are based on the Exceptional Generalised Geometry (EGG) formalism developed in \cite{EGG}.

The $d.o.f.$ which lead to the scalars after reductions are combined to form a unique geometric object, the {\it generalised vielbein}. This object reduces to the $\hat V^{\mathbb A}{}_{\mathbb M}$ of the previous section if the so called {\it section condition} is imposed (see below). Also fermions fit nicely in this formulation, but they lack of a geometrical interpretation. In addition, local transformations, including diffeomorphisms and gauge transformations are unified in a unique generalised Lie derivative, while local Lorentz symmetry is promoted to the maximal compact subgroup of $E_{7(7)}$, $SU(8)$.

The great potentiality of these frameworks rests in that truncations with non trivial curved manifolds, like spheres and hyperbolic spaces can be described as generalised Scherk-Schwarz reductions, $i.e.$ the nonlinear uplift formulas for the metric, 3-form and dual 6-form turn out to be linear with the new rearrange of $d.o.f.$
\begin{equation}
	\label{genviel}
	V^{\mathbb A}{}_{\mathbb M}(x^\mu,Y^{\mathbb N}) = {\cal V}^{\mathbb A}{}_{\mathbb B}(x^\mu) E^{\mathbb B}{}_{\mathbb M}(Y^{\mathbb N}), 
\end{equation}
where ${\cal V}(x)$ encodes the coset representative of the $E_{7(7)}/SU(8)$ scalar manifold and $E\in {\mathbb R}^+\times E_{7(7)}/SU(8)$ denotes the generalised vielbein of the extended internal 56D space
\begin{eqnarray}
E_{\mathbb A}{}^{\mathbb M}= e^{-\rho}~ {\cal E}_{\mathbb A}{}^{\mathbb M}.   
\end{eqnarray}

${\cal E}_{\mathbb A}{}^{\mathbb M}$ can be seen as a vector, transforming under the generalised diffeomorphisms encoding both diffeomorphisms and gauge transformations of the 3 and dual 6-form in only one parameter $\xi^{\mathbb M}$
\begin{eqnarray}
\delta_\xi V^{\mathbb M}=\xi^{\mathbb P} \partial_{\mathbb P} V^{\mathbb M}
-12 P_{(adj)}{}^{\mathbb M}{}_{\mathbb N}{}^{\mathbb P}{}_{\mathbb Q} \partial_{\mathbb P}\xi^{\mathbb Q} V^{\mathbb N} 
+ \frac\omega2 \partial_{\mathbb P}\xi^{\mathbb P} V^{\mathbb M}~.\label{GenDiff}
\end{eqnarray}

The first two terms in (\ref{GenDiff}) guarantee the preservation of the $E_7$ structure and the last term comes from the ${\mathbb R}^+$ transformations, the weight $\omega$ is 1 for $E$ but vanishes for the $E_7/SU(8)$ bein ${\cal E}$ and $P_{(adj)}$ denotes the projector on the adjoint representation (see (\ref{padj})). The closure of the algebra is guaranteed\footnote{As was discussed in \cite{EG} it is a sufficient but not necessary condition and could be relaxed in special configurations.} if all fields satisfy the $E_7$ covariant constraints known as {\it section conditions} 
\begin{eqnarray}
\Omega^{{\mathbb M} {\mathbb N}}\partial_{{\mathbb M}}{\cal A}~\partial_{\mathbb N}{\cal B}=0,~~
[t_\alpha]^{{\mathbb M}{\mathbb N}}\partial_{{\mathbb M}}{\cal A}~\partial_{\mathbb N}{\cal B}=0,~~~[t_\alpha]^{{\mathbb M}{\mathbb N}}\partial_{{\mathbb M}}\partial_{\mathbb N}{\cal A}=0~,\label{SecCond}
\end{eqnarray}
where ${\cal A},{\cal B}$ denote arbitrary fields or gauge parameters and $t_\alpha$ the generators of $\mathfrak{e}_{7(7)}$ in the fundamental representation. These constraints are so strong that imply that fields have a dependence in at most 7 coordinates and EFT turns out to be just an appropriate rewritten of 11D supergravity.
 
Similar to ordinary Scherk-Schwarz reductions where the Lie bracket of the twist matrix leads to the structure constants of the gauge group in the effective theory, here the generalised Lie derivative of the internal generalised vielbein, $E$ generates to the embedding tensor of the truncated 4D supergravity
\begin{equation}
	L_{E_{\mathbb A}} E_{\mathbb B}\equiv  \delta_{E_{\mathbb A}} E_{\mathbb B} = F_{{\mathbb A}{\mathbb B}}{}^{\mathbb C} E_{\mathbb C}\label{LieDerE}
\end{equation}
Equation (\ref{GenDiff}) leads to
\begin{eqnarray}
&&\;\;\;\;\;\;\;\;\;\;\;\;\;\;\;\;\;\;\;\;\;
F_{{\mathbb A} {\mathbb B}}{}^{{\mathbb C}}=X_{{\mathbb A} {\mathbb B}}{}^{{\mathbb C}}+D_{{\mathbb A} {\mathbb B}}{}^{{\mathbb C}},\\
&&X_{{\mathbb A} {\mathbb B}}{}^{{\mathbb C}}=\Theta_{\mathbb A}{}^{\alpha}\left[t_\alpha\right]_{\mathbb B}{}^{\mathbb C}~,~~~~~~\
\Theta_{\mathbb A}{}^{\alpha}=7P_{(912)}{}_{\mathbb A}{}^{\alpha}{},^{B}{}_{\beta}\tilde\Omega_{\mathbb B}^{\beta},\label{Xtensor}\\
D_{{\mathbb A} {\mathbb B}}{}^{{\mathbb C}}&=&-\vartheta_{\mathbb A}\delta_{\mathbb B}^{\mathbb C}+ 8~P_{(adj)}{}^{\mathbb C}{}_{\mathbb B}{}^{\mathbb D}{}_{\mathbb A}\vartheta_{\mathbb D}~,~~~\vartheta_{\mathbb A}=-\frac12 \left(\tilde\Omega_{{\mathbb B}{\mathbb A}}{}^{\mathbb B}-3\partial_{\mathbb A}\rho\right)~,
\end{eqnarray}
wherein $X_{{\mathbb A} {\mathbb B}}{}^{{\mathbb C}}$ and $D_{{\mathbb A} {\mathbb B}}{}^{{\mathbb C}}$ are the projections in the ${\bf 912}$ and ${\bf 56}$ representations, generated respectively by
\begin{eqnarray}\label{padj}
P_{(912)A}{}^{\alpha},{}^{B}{}_{\beta}=\frac17\delta^\alpha_\beta \delta_A^B-\frac{12}{7}
\left[ t_\beta t^\alpha\right]_A{}^B +\frac{4}{7}\left[t^\alpha t_\beta\right]_A{}^B\;,\;\;\;\;\;\;\;\;\;\;
P_{(adj)}{}^{A}{}_{B}{}^{C}{}_{D}=\left[ t_\alpha\right]_B{}^A~\left[ t^\alpha\right]_D{}^C\;,
\end{eqnarray}
and $\tilde \Omega_{{\mathbb A} {\mathbb B}}{}^{{\mathbb C}}$ is the flat index Weitzenb\"ock connection of the $E_7$ piece, 
\begin{eqnarray}
\tilde\Omega_{{\mathbb A}{\mathbb B}}{}^{{\mathbb C}}=e^{-\rho}~ {\cal E}_{\mathbb A}{}^{\mathbb M}~ 
{\cal E}_{\mathbb B}{}^{\mathbb N}~ \partial_{\mathbb M} {\cal E}^{\mathbb C}{}_{\mathbb N}~,\label{Weitzenbock}
\end{eqnarray}
If one considers situations with vanishing $\vartheta$, the allowed fluxes, $X$ satisfy the 4D maximal supergravity relations of the embedding tensor 
\begin{eqnarray}
P_{(adj)}{}^{\mathbb C}{}_{\mathbb B}{}^{\mathbb D}{}_{\mathbb E}~X_{\mathbb A\mathbb D}{}^{\mathbb E}= X_{\mathbb A\mathbb B}{}^{\mathbb C}~ ,
~~~X_{\mathbb A[\mathbb B\mathbb C]}=X_{\mathbb A\mathbb B}{}^{\mathbb B}=X_{(\mathbb A\mathbb B\mathbb C)}=X_{\mathbb B\mathbb A}{}^{\mathbb B}= 0.
\end{eqnarray}

Hence this gives a top down framework to uplift gauged supergravities. Every maximally supersymmetric deformation is characterized by an embedding tensor and the problem of finding a higher dimensional origin to a given theory reduces to integrate the linear differential equations (\ref{LieDerE}). 
 
This approach was applied to give a no go theorem for the generation of dyonic $SO(8)$ gauged supergravities in some special sectors of the parameter space \cite{Baron:2014yua} and a complete proof was proposed very recently by using the parent EGG framework in \cite{Lee:2015xga}.

In addition, the EFT approach was successfully implemented to uplift non compact electric gauged supergravities. The solution to (\ref{LieDerE}) for $SO(p,q)$ and $CSO(p,q,r)$ gaugings was found in \cite{Hohm:2014qga} simultaneously with the bottom up approach for $SO(p,q)$ in \cite{Baron:2014bya}. It is worth mentioning that even though $loc.cit.$ solved the non compact gaugings by using the framework of the first section, the authors explicitly verified that the generalised vielbein is a solution to the differential equations of this section. The analogous verification in the compact $SO(8)$ gaugings was performed in \cite{Lee:2014mla}.

Let us conclude this section by further extending the map between both frameworks. The conformal factor $e^{-\rho}$ in these truncations is simply given by $\Delta^{\frac12}$ and the {\it generalised vielbein}, $\hat V^{\mathbb A}{}_{\mathbb M}$ of section 2 can be parametrized in the {\bf 56} $\rightarrow$ {\bf 21}+{\bf 7}+{\bf 21}'+{\bf 7}' decomposition as
\begin{eqnarray}
\left(\begin{matrix}
\Delta^{\frac12}\,e_{a}{}^{[m}\,e_{b}{}^{n]}              &                   0                                          &  
             0                                            &    \frac{\sqrt2}{2} \Delta^{\frac12}\,e_{a}{}^p\,e_{b}{}^q A_{mpq}   \cr
2\Delta^{\frac12}\,e_{\underline{a}}{}^p S_-^{mn}{}_p     &    \Delta^{-\frac12}\,e_{a}{}^m                              &  
\sqrt2\Delta^{-\frac12}\,e_{a}{}^p A_{mnp}                &    \Delta^{\frac12}\,e_{a}{}^p S_{mp}                       \cr
-\Delta^{\frac12}\,e^{a}{}_p\,e^{b}{}_q S^{mnpq}          &                   0                                          &   
\Delta^{-\frac12}\,e^{a}{}_{[m} \,e^{b}{}_{n]}            &    -\Delta^{\frac12}\,e^{a}{}_p\,e^{b}{}_q S_+^{pq}{}_m     \cr
             0                                            &                   0                                          & 
						 0                                            &    \Delta^{\frac12}\,e^{a}{}_m                              \cr
\end{matrix}\right),\label{Vmatrix}
\end{eqnarray}
with $e^{a}{}_m$ and $A_{mnp}$ representing the internal components of the vielbein and 3 form of the 11-dimensional supergravity, so that they have implicit dependence on all 11 coordinates. 
The other tensors appearing above are defined as
\begin{eqnarray}
S_{\pm}^{mn}{}_{p}&=&3\sqrt2\stackrel{\circ}\epsilon{}^{mn q_1\dots q_5}
\frac{1}{6!}\left(\tilde A_{p q_1\dots q_5}\pm 5\sqrt2  A_{p q_1 q_2}A_{q_3 q_4 q_5}\right),\\[2mm]
S_{mn}&=&-\frac{1}{5!}\stackrel{\circ}\epsilon{}^{q_1\dots q_7}A_{m q_1 q_2}\left(\tilde A_{n q_3\dots q_7}
-\frac{\sqrt2}{3} 5A_{n q_3 q_4}A_{q_5 q_6 q_7}\right),\\[2mm]
S^{mnpq}&=&\frac{\sqrt2}{3!\,2}\stackrel{\circ}\epsilon{}^{mnpq r_1 r_2 r_3} A_{r_1 r_2 r_3},
\end{eqnarray}
where $\tilde A$ is the dual six form, defined in (\ref{6form}) and, $\stackrel{\circ}\epsilon_{q_1\dots q_7}$ and $\epsilon_{\hat M_1\dots \hat M_{11}}$ denote, respectively the volume form of the Lorentzian hyperboloid (seven sphere in the $p=8$ situation) and the one of the 11 dimensional manifold.

Strictly speaking (\ref{Vmatrix}) is not of the form (\ref{genviel}). Indeed, after tedious but straightforward computations one can plug the non linear ansatz for the metric, 3-form and dual 6-form to factorize $\hat V$ as
\begin{eqnarray}
{\hat V}_{\mathbb A}{}^{\mathbb M}(x,y)={\cal W}_{\mathbb A}{}^{\mathbb B}(y) 
{\cal V}_{\mathbb B}{}^{\mathbb C}(x) {\cal E}_{\mathbb C}{}^{\mathbb M}(y).
\end{eqnarray}

Nevertheless it can be shown that ${\cal W}_{\mathbb A}{}^{\mathbb B}(y)$ is pure gauge and so it can be neglected. 

\section{Conclusions and Outlook}
Supergravity theories are generally believed to describe some low-energy approximation of string theory models. There is, however, a huge landscape of 4-dimensional supergravity models whose higher-dimensional origin is not clear yet. This is already true for the most
constrained scenario of maximal supergravities. Here we discuss the uplift of these theories in the restricted case of electric gaugings using the two main approaches available in the literature, a bottom up approach that uses supersymmetry transformations as a guiding principle to construct the uplift formulae and a more speculative frame that uses Exceptional Geometries and a generalised form of the Scherk-Schwarz reduction procedure, which offers certain advantages. For instance, the internal and the space-time coordinate dependence factorizes linearly and the search of a higher dimensional origin (if it exist at all) of a particular gauged supergravity theory translates in the resolution of a differential equation.

Generalised Scherk-Schwarz reductions have also been shown to be a powerful technique in other situations. For instance in the context of the closely related Double Field Theory, this approach was successfully applied to describe the electric sector of half maximal supergravity \cite{Aldazabal:2011nj}. This special reduction procedure was successfully applied in DFT in situations where the section condition can be relaxed leading to genuinely non geometric reductions, giving a higher dimensional interpretation of dyonic $SO(4)$ gaugings in 7D \cite{Dibitetto:2012rk}. 

It is worth mentioning that \cite{Lee:2015xga} stated that such a dyonic gaugings cannot be generated in maximal supergravity without violating the section condition and indeed proposed a solution whose physical interpretation is still not clear.

\end{document}